\documentclass[12pt]{iopart}
\usepackage{graphicx}
\usepackage{iopams}
 \usepackage{cite}
\begin{document}

\title{Integrable order parameter dynamics of globally coupled oscillators}

\author{G~M~Pritula, V~I~Prytula, O~V~Usatenko}

\address{O Ya Usikov Institute for Radiophysics and
Electronics, National Academy of Sciences of Ukraine, 12 Proskura Street, 61085
Kharkov, Ukraine}
\ead{pritula.galina@gmail.com}

\begin{abstract}

We study the nonlinear dynamics of globally coupled nonidentical oscillators in the framework of two order parameter (mean field and amplitude-frequency correlator) reduction. The main result of the paper is the exact solution of a corresponding nonlinear system on a two-dimensional invariant manifold. We present a complete classification of phase portraits and bifurcations, obtain explicit expressions for  invariant manifolds (a limit cycle among them) and derive analytical solutions for arbitrary initial data and different regimes.

\end{abstract}

%%%%%%%%%%%%%%%%%%%%%%%%%%%%%%%%%%%%%%%%%%%%%%%%%%%%%%%%%%%%%%%%%%%%%%%%%%%%%%%%
\section{Introduction}
%%%%%%%%%%%%%%%%%%%%%%%%%%%%%%%%%%%%%%%%%%%%%%%%%%%%%%%%%%%%%%%%%%%%%%%%%%%%%%%%

Frequency synchronization of nonlinear interacting oscillators is one of the most common phenomena in nature. The phenomenon discovered by Huygens in the 17th century was found out later in astronomy, chemistry and electronics, and manifests itself in various technical devices, environmental systems and nuclear physics. Today, the direction of research has shifted to studying complex nonlinear dynamical systems with global interaction such as the cooperative dynamics of laser arrays and arrays of Josephson junctions; different forms of collective behavior of insects, birds, animals and human communities; adjustment of heart rate and respiration, synchronization of electrical potentials of brain neurons and so on. Recently new phenomena such as, for example, the partial synchronization, clustering and chimera states have been revealed. There is a large and rapidly increasing amount of literature on the collective oscillatory behavior among which one can find already classic books by Kuramoto \cite{K}, Strogatz \cite{StrSync,StrND}, and Pikovsky et al. \cite{PRK}, the review by Asebron et al \cite{As}, and the recent brief but very informative review by Pikovsky and Rosemblum \cite{PR} on the progress and perspectives of studies of globally coupled oscillators. 

Despite the apparent diversity of the phenomena emerging in ensembles of interacting nonlinear oscillators, many of them can be modeled by a system of equations based on the Stuart-Landau one, which is closely associated with the phenomenon of the supercritical Hopf bifurcation. Y. Kuramoto was the first to describe synchronization transition in oscillatory reaction-diffusion systems \cite{K75,K} reducing the system of globally coupled Stuart-Landau equations
to the system of the coupled phase oscillators known now  as the famous Kuramoto model.
Later the amplitude-phase dynamics was reinstated in a series of papers \cite{MetMirollo,MetStr,DMO,DMOM}. The mixed amplitude-phase dynamics significantly expanded a range of phenomena elucidated with the Kuramoto model. Depending on the ratio between the amount of interaction and the deviation of natural oscillator frequencies, new regimes of collective behavior such as amplitude death, quenched amplitude, oscillations with large amplitude, collective chaos and intermediate mixed states can arise.
 
The system of equations describing the dynamics of globally coupled oscillators is a nonlinear system, and therefore, in general, is nonintegrable. However, under certain assumptions the system can be simplified and reduced to a solvable one. This occurs, for example, when the interaction is strong enough and the system of equations for globally coupled nonidentical Stuart-Landau oscillators is reduced within the self-consistent field approach to a system of equations for two macroscopic order parameters, the mean field and the amplitude-frequency correlator called the shape parameter. In our work we integrate this  nonlinear system on an invariant manifold, present the complete classification of phase portraits and bifurcations, obtain exact expressions for the invariant manifolds (a limit cycle among them) and derive analytical solutions for arbitrary initial data in different regimes.
 
The paper is organized as follows. In section \ref{OrdPar} we present the dynamic equations for the self-consistent field and reduce them to the solvable problem. section \ref{Diag&Man} provides a phase diagram of system (\ref{mainsystem}) for the order parameters  and analytical expressions for the invariant manifolds. In section \ref{Analit} we obtain the exact analytical solution of system  and in section \ref{Results} we discuss the results.

\section{Two order parameter approximation}\label{OrdPar}

The model we are dealing with is a system of $N$ nonidentical nonlinear 
Stuart-Landau oscillators with a global dissipative linear coupling \cite{MetMirollo}:
\begin{equation}
  \frac{dz_j}{dt} =
  ( 1 + i{\omega}_j - |z_j|^2)z_j
  + K {1 \over N} \sum_{l=1}^N (z_l - z_j),
\label{Hforms}
\end{equation}
where $j=1,...N$; $z_j(t)$ is a position of the $j$-th oscillator in the complex
plane, $K \ge 0$ is the coupling strength and randomly distributed natural
frequencies ${\omega}_j$ are characterized by the average frequency ${\omega}_0$ and
the standard deviation ${\sigma}$. When the coupling is absent ($K=0$)  system
(\ref{Hforms}) describes the dynamics of noninteracting oscillators with
a stable circular limit cycles $|z_j|=1$.

In \cite{DMO} system (\ref{Hforms}) was studied in the framework of two order parameter approximation
(for more about the order parameter expansion method, see  \cite{DMOM, DMOHM04, DMOHM05, GomesToral06, KominToral10, KominToral_PR10}):
\begin{equation}
Z={1 \over N}\sum_{l=1}^N z_l
\label{Z}
\end{equation}
and
\begin{equation}
  W = {1 \over N}\sum_{l=1}^N
  ({\omega}_l - {\omega}_0)(z_l - Z).
\label{W}
\end{equation}
Here $Z$ is a coordinate of centroid, the ``center of mass" of the ensemble, $W$ is the so-called shape parameter, the cross-correlator of amplitude and phase variations.
For a sufficiently strong coupling and narrow frequency distributions when 
the distance $|Z-z_j|$ can be considered small, in \cite{DMO} system of equations (\ref{Hforms}) 
up to terms of the first order in
$|Z-z_j|$ 
was reduced to a closed system of equations for the macroscopic order parameters:
\begin{equation}
  \left\{
  \begin{array}{lcr}
    \displaystyle{ \frac{dZ}{dt}} =
    ( 1 + i{\omega}_0 - |Z|^2)Z + iW,
  \\[3mm]
    \displaystyle{\frac{dW}{dt}} =
    i{\sigma}^2Z + (1 - K - 2|Z|^2 + i {\omega}_0)W - Z^2W^*.
  \end{array}
  \right.
\label{system1}
\end{equation}
This system was derived by combining the averaging of original equations (\ref{Hforms})  with a perturbative approach based on the smallness of individual oscillator displacements from the position of centroid $|Z-z_j|$ compared with $|Z|$ along with neglecting higher-order correlators, which allowed one to omit the terms like $\left<|z_j-Z|^2\right>$ as well as $\left<(\omega_j-\omega_0)^2 (z_j-Z)\right>$. Though this derivation, formally speaking, is valid only for $|\omega_j-\omega_0|<< \omega_0$, the comparison of the results of  \cite{DMO} with numerical simulations for original system (1) \cite{MetMirollo} demonstrated that (\ref{system1}) reproduces qualitative features of the population dynamics in rather a wide region of $\sigma$ and $K$ (see, e.g., \cite{As,GomesToral06}).

The resulting system allows us to take into consideration the effect of the mismatch of oscillator phases  on the dynamics of the centroid. Presenting solutions of the system in the form $Z=re^{i({\omega_0}t+\theta)}$ and $W=we^{i({\omega_0}t+\theta+\Delta)}$ with real $r$ and $w$ and separating the real and 
imaginary parts, in a reference frame rotating at frequency $\omega_0$, we obtain
\begin{equation}
	\left\{  
	\begin{array}{lcr}
		 \dot{r}=(1-{{r}^{2}})r-w\sin \Delta , 
		\\ [3mm]
			\dot{w}={{\sigma }^{2}}r\sin \Delta +(1-K)w-
			(1+2\cos^2 {\Delta} ){{r}^{2}}w, 
		\\ [3mm]
			\dot{\theta }=\displaystyle{\frac{w}{r}}\cos \Delta , 
		\\ [3mm]
			\dot{\Delta }=\left( \displaystyle{\frac{r}{w}}{{\sigma }^{2}}+2{{r}^
				{2}}\sin \Delta -\frac{w}{r} \right)\cos \Delta  \\ 
	\end{array} 
  \right.
\label{systemRI}
\end{equation}
where dots stand for the derivatives with respect to the time. 
As seen from (\ref{systemRI}), the plane $\Delta=\pi/2$ is an invariant manifold, i.e. 
a manifold embedded in a phase space with the property that the trajectories that start out in the manifold remain in it.
A comprehensive analysis of stability of this orthogonal manifold is cumbersome and out of the scope of the paper. It can be shown that this manifold is attracting   
in the limit of infinitely large $K$, when the 
solutions of (\ref{systemRI}) 
obtain the form of the following expressions with the corresponding 
asymptotics for $t\rightarrow \infty$:
\begin{equation}
\begin{array}{lcl}
   w&\approx&{{w}_{0}}{{e}^{-Kt}}\to 0,
	\\[3mm] 
  r&\approx&\displaystyle\pm {\frac{r_0}{\sqrt{r_{0}^{2}+(1-r_{0}^{2}){e}^
{-2t}}}}\to \pm 1, 
	\\[7mm]
  \theta &\to &{{\theta }_{0}}=const, 
	\\[2mm]
  \Delta &\approx&\displaystyle\sin^{-1} \left( \tanh\,\left(\frac{{\sigma}^{2}r_{0}}{w_{0}}\int\limits_{{{t}
_{0}}}^{t}{{{e}^{Kt}}dt} \right) \right)\to \frac{\pi }{2}. 
\end{array}
 \label{attractivM}
\end{equation}

The evaluation for $\Delta$ in equation (\ref{attractivM}) can be made with the help of the new variable $F=\tanh^{-1}(\sin{\Delta})$ in the last expression of system (\ref{systemRI}). After this change of variable, substituting the first and second expressions of (\ref{attractivM}) into the obtained equation for $F$ and keeping the largest terms in its right hand side, we get an easily integrated equation for our new variable: $\dot{F}={\sigma}^{2}r_{0}w_{0}^{-1}{{e}^{Kt}}$, which gives the last expression of set (\ref{attractivM}) and indicates the attractivity of the manifold under the initial conditions $w_0$ and $r_0$ of the same signs, $r_0/w_0>0$. 

Some insight on the physical interpretation of this invariant orthogonal manifold can be gained from the analysis of the energy balance for the system. It is possible to show that the orthogonality condition $\Delta=\pi /2$ or $\mbox{Re}\,ZW^*=0$ is equivalent, to some extent, to the condition for conservation of energy of the system up to terms that were taken into consideration for deriving system of equations (\ref{system1}).

In the projection  onto the plane $\Delta=\pi/2$, system (\ref{systemRI}) is reduced to the following:
\begin{equation}   
  \left\{
  \begin{array}{lcl}
     \dot r & = & \left( 1 - r^{2} \right)r - w,
  \\[3mm]
     \dot w & = & \sigma^{2} r + \left( 1 - K - r^{2} \right) w.
  \end{array}
  \right.
\label{mainsystem}
\end{equation}

Equations (\ref{mainsystem}) represent the dynamics of system (\ref{system1}) on its invariant
manifold. The authors of \cite{DMO} used these equations to extract some
analytical expressions, such as the values of the amplitudes in the fixed points and
the quenching radius, for the qualitative and numerical description of the
collective properties of the model. In our work we show that despite the nonlinearity system (\ref{mainsystem}) can be integrated exactly.
The main result of our work is an  analytical solution of this system of nonlinear differential equations.

%%%%%%%%%%%%%%%%%%%%%%%%%%%%%%%%%%%%%%%%%%%%%%%%%%%%%%%%%%%%%%%%%%%%%%%%%%%%%%%%
\section{Phase diagram and invariant manifolds}\label{Diag&Man}
%%%%%%%%%%%%%%%%%%%%%%%%%%%%%%%%%%%%%%%%%%%%%%%%%%%%%%%%%%%%%%%%%%%%%%%%%%%%%%%%

\begin{figure}[h]%  
\begin{center}               
  \includegraphics[width=100mm]{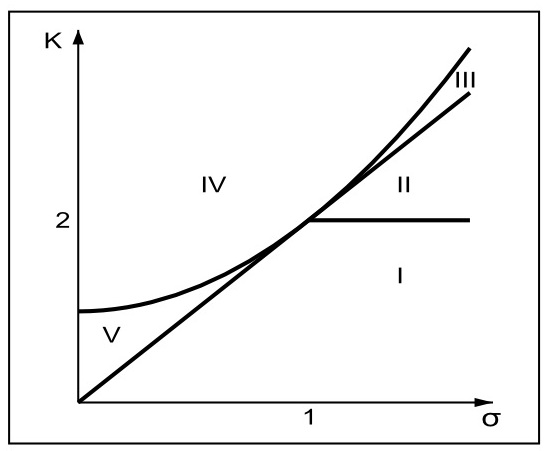}
\end{center}
  \caption{The phase diagram  for system (\ref{mainsystem}).}
  \label{fig:PhaseDiagram}
\end{figure}

The dynamic behavior of the system under consideration depends essentially on 
the relationship between the coupling constant $K$ and the standard deviation of random oscillator frequencies $\sigma$.
It can be shown that the $(K,\sigma)$-plane is divided
into five regions with different number and character of fixed
points and hence with a different behavior of the system. The
phase diagram is depicted in figure \ref{fig:PhaseDiagram}. 
The regions of different dynamic behavior are separated by
the lines $K=2\sigma$ and $K=2$ and the
parabola $K={\sigma}^2+1$.

Since the main purpose of the work is an exact solution of equations (\ref{mainsystem}), we omit the description of the standard qualitative analysis of the behavior of the system in the phase plane confining ourselves to the corresponding phase portraits (figures \ref{fig:Fig2}-\ref{fig:Fig6}), and dwell on obtaining analytical expressions for the invariant manifolds.

To simplify analysis of system
(\ref{mainsystem}) 
and make results more intelligible we introduce the new variable
\begin{equation}
 s=w+(\xi +\eta -1)r,
  \label{newvarS}
\end{equation}
where the new parameters $\xi$ and $\eta$ are given by
\begin{equation}
\begin{array}{lcl}
     \xi & = & 1-K/2+\sqrt{(K/2)^2-{\sigma}^2},
  \\[3mm]
     \eta & = & 1-K/2-\sqrt{(K/2)^2-{\sigma}^2}.
  \end{array}
  \label{zeta}
\end{equation}
Thus, we rewrite our system as
\begin{equation}
  \left\{
  \begin{array}{lcl}
     \dot r & = & \left( \xi + \eta \right)r - s -r^3,
  \\[3mm]
     \dot s & = & \xi \eta  r - r^2 s
  \end{array}
  \right.
\label{magicmainsyst}
\end{equation}
and will hereafter analyze the problem in the terms of
$(r,s)$-variables.

 Let us now consider the following quantity
$\Gamma=\Gamma(r,s)$:
\begin{equation}
  \Gamma = a r^2 + b s^2 - 1.
  \label{gama1}
\end{equation}
By differentiating it and substituting the expressions for $\dot
r$ and $\dot s$ from system (\ref{magicmainsyst}) we get
\begin{equation}
  \dot\Gamma =
  - 2 r^2 [a r^2 + b s^2 - b \xi \eta (\xi + \eta)]
  - 2(a - b \xi \eta) r s.
  \label{gamader1}
\end{equation}
Choosing the parameters $a$ and $b$ as
\begin{equation}
  a = \frac {1}{\xi + \eta},  
\qquad \qquad
  b = \frac{1}{\xi \eta (\xi + \eta)}
\end{equation}
one can conclude that the evolution of  quantity $\Gamma$ is governed by the
simple equation:
\begin{equation}
  \dot \Gamma = - 2 r^2 \Gamma.
  \label{gamader2}
\end{equation}
Thus, if at the initial moment of time the quantity $\Gamma_0 = \Gamma(t=0) = 0$,
then it will be zero for all the values of $t$, which means that the
second order curve
\begin{equation}
  \frac{r^2}{\xi + \eta} + \frac{s^2}{\xi \eta (\xi + \eta)} = 1
  \label{Cycle}
\end{equation}
is an invariant manifold.

When the parameters $K$ and $\sigma$ belong to region I of the
phase diagram the coefficients of (\ref{Cycle}) are positive and
this second order curve, which is a stable limit cycle, is an
ellipse. In such a way we have derived the explicit analytical
expression describing the form of the limit cycle of our system (\ref{mainsystem}).

As we move up along the line $\sigma =  const> 1$ (figure \ref{fig:PhaseDiagram}),
the axes of the ellipse become smaller and smaller until at $K =
2$ the limit cycle degenerates into the center, the $(0,0)$-point,
which further transforms into the stable focus of region II.
This phenomenon is the well-known Hopf bifurcation \cite{Kuzn} and
we would like to note that our system allows analytical description of this
bifurcation, which is often not the case.

The line $K=2\sigma > 2$ separates the regions with different types of the stable fixed point $(0,0)$: the stable focus of region II transforms into the stable node of region III. This line is actually not a bifurcation border from the view point of the dynamics of  system (\ref{mainsystem}), because the topology of its phase space is not changed.  

One more well-known bifurcation, the pitchfork one, takes place at
the boundary between regions III and IV: $K=\sigma^2+1 >2$,
where the stable node of region II looses its
stability  and transforms into the saddle (0,0) and two stable
nodes $(\pm{\sqrt{\xi}},\mp{\eta \sqrt{\xi}})$ in region IV.
In this region $\xi + \eta < 0$, but $\xi \eta (\xi + \eta)> 0$ and
expression (\ref{Cycle}) describes a hyperbola.

Immediately after crossing the border between regions IV and V, in region V the saddle point $(0,0)$  transforms into the unstable node $(0,0)$ originating two saddles $(\pm{\sqrt{\eta}},\pm{\xi\sqrt{\eta}})$ while the stable nodes $(\pm{\sqrt{\xi}},\pm{\eta \sqrt{\xi}})$ remain. In region V both $\xi$ and $\eta$ are positive and line
(\ref{Cycle}) is again an ellipse. However, contrary to region I, 
 this ellipse now consists of four
separatrices which start at the saddles $(\pm{\sqrt{\eta}},\pm{\xi
\sqrt{\eta}})$ and end at the stable nodes
$(\pm{\sqrt{\xi}},\pm{\eta \sqrt{\xi}})$. Thus, our simple
calculations enabled us to establish the form of the separatrices
(the parts of the ellipse). 

While approaching to the boundary separating regions V and I, $K=2\sigma < 2$, the angle between the straight-line invariant manifolds $s=\xi r$ and $s=\eta r$ decreases until, at
the boundary, they merge into one
unstable at the origin 
manifold $s=(1-\sigma)r$ along with the merging of the saddle and node points. 
This creates saddle-node points at the crossing with the ellipse. 
Further in region I the  origin transforms into the unstable focus, the linear manifold disappears and the elliptic manifold becomes the limit cycle. At the border between regions IV and V the pairwise saddle-node or heteroclinic fold bifurcation \cite{Kuzn} occurs.

Near the codimension-two point $\sigma=1,K=2$, where all the regions of the phase diagram of our two-dimensional system meet and the bifurcation borders cross, in  high dimensional system  (\ref{Hforms}) different dynamical regimes of the adjacent regions can mix.
As was shown by the numerical study \cite {MetMirollo} in the phase diagram of the original system (\ref{Hforms}) near this point  there is a region of unsteady dynamical behavior.

%%%%%%%%%%%%%%%%%%%%%%%%%%%%%%%%%%%%%%%%%%%%%%%%%%%%%%%%%%%%%%%%%%%%%%%%%%%%%%%%
\section{Analytical results}\label{Analit}
%%%%%%%%%%%%%%%%%%%%%%%%%%%%%%%%%%%%%%%%%%%%%%%%%%%%%%%%%%%%%%%%%%%%%%%%%%%%%%%%

The main aim of this paper is stated as being the exact
integration of system (\ref{mainsystem}) or, with the use of the
variables $r$ and $s$, system (\ref{magicmainsyst}). The most
straightforward way to do so consists of two steps. First, from
(\ref{magicmainsyst}) we can easy derive that the quantity $s/r$
satisfies a simple equation
\begin{equation}
  \frac{d}{dt} \left( \frac{s}{r}\right) =
  \left( \frac{s}{r} -  \xi \right)
  \left( \frac{s}{r} - \eta \right).
  \label{simplequation}
\end{equation}
Looking at this equation one can immediately note that there are two stationary
solutions $s/r=const$:
\begin{equation}
  \frac{s}{r}= \xi,
\qquad \qquad
  \frac{s}{r}= \eta
\end{equation}
that correspond to the linear invariant manifolds represented in figures \ref{fig:Fig4}--\ref{fig:Fig6}
(in the cases I and II parameters $\xi$ and $\eta$ are complex and these
solutions 
do not any physical meaning).

Integrating equation (\ref{simplequation}) we can find that our
variables $r$ and $s$ are related by
\begin{equation}
  \frac{s}{r} = \frac {\eta E^2-\xi F_0 }{E^2 - F_0}
  \label{solsimplequation}
\end{equation}
where
\begin{equation}
  F_0 = \frac {s_0 -\eta r_0}{s_0 -\xi r_0}
\label{F0}
\end{equation}
and
\begin{equation}
  E^2= \exp{(\xi -\eta)t},
\label{E^2}
\end{equation}
$s_0, r_0$ are initial conditions of the
problem.

The fact that the ratio $s/r$ satisfies the closed equation
(\ref{simplequation}) which can be integrated explicitly makes it
possible to reduce the order of our initial system and complete its
solution. In some sense equation (\ref{solsimplequation}) plays the role of
integral of motion (energy) in conservative systems.

To obtain $r$ and $s$ themselves we will present them in the
following form:
\begin{eqnarray}
  r & = & u \left(\alpha E +\beta E^{-1}\right),
 \label{commonform1}
\\
   s & =& u \left(\alpha \eta E +\beta \xi E^{-1}\right).
   \label{commonform2}
\end{eqnarray}
Here $u=u(t)$ is a function that
has to be found, the constants $\alpha$ and $\beta$, related by
$\beta/\alpha= - F_0$, are determined by the initial conditions of the
problem.
This representation is a direct consequence of (\ref{solsimplequation})
which is rewritten in a more symmetric form convenient for the further analysis.

Substituting (\ref{commonform1}) and (\ref{commonform2}) into
equation (\ref{magicmainsyst}) and performing some algebraic
transformations we arrive at the equation for the function $u(t)$:
\begin{equation}
  \displaystyle{\frac{du}{dt}
  = \frac{\xi+\eta}{2}u
  - {\left(\alpha E +\beta E^{-1}\right)}^2u^3}.
  \label{equatU}
\end{equation}
This equation is the so-called Abel equation of the first kind \cite{Zwil}.
Using the substitution
\begin{equation}
 u=v^{-1/2}
 \label{substitution}
\end{equation}
one can convert (\ref{equatU}) into the linear equation
\begin{equation}
  \displaystyle{\frac{d}{dt}v} + (\xi+\eta)v
  = 2{\left(\alpha E+\beta E^{-1}\right)^2}.
  \label{equatV}
\end{equation}
The solution of this equation is
\begin{equation}
  v=v_* e^{-(\xi+\eta)t}+\displaystyle{\frac{\alpha^2}{\xi}
  e^{(\xi-\eta)t}
  +\frac{\beta^2}{\eta}  e^{(\eta-\xi)t}
  +\frac{4\alpha \beta}{\xi+\eta}},
  \label{solutionV}
\end{equation}
where the constant $v_*$ should be determined from the initial
conditions. That gives
\begin{equation}
  v_* =1
  -\displaystyle{\frac{\alpha^2}{\xi}
  -\frac{\beta^2}{\eta}
  -\frac{4\alpha \beta}{\xi+\eta}}
  \label{Vstar}
\end{equation}
and
\begin{equation}
  \alpha=\displaystyle{\frac{s_0-\xi r_0}
 {\eta-\xi}},
  \label{alpha}
\end{equation}
\begin{equation}
  \beta=-\displaystyle{\frac{s_0-\eta r_0}
 {\eta-\xi}}.
  \label{beta}
\end{equation}
Here we assumed without breaking the generality of the solution that
$u(t=0)={v_0}^{-1/2}=1$ (for a system of two ordinary differential 
equations of the first
order we need just two arbitrary constants).

Thus, we have integrated our nonlinear problem and expressions
(\ref{commonform1}),
(\ref{commonform2}) together with (\ref{solutionV})-(\ref{beta}) give us 
its complete analytical solution.
As one can see, the obtained solution has  quite a compact form
convenient for analysis. 

%%%%%%%%%%%%%%%%%%%%%%%%%%%%%%%%%%%%%%%%%%%%%%%%%%%%%%%%%%%%%%%%%%%%%%%%%%%%%%%%
\section{Analysis of the results}\label{Results}
%%%%%%%%%%%%%%%%%%%%%%%%%%%%%%%%%%%%%%%%%%%%%%%%%%%%%%%%%%%%%%%%%%%%%%%%%%%%%%%%

%

In this section we continue
the description of the system behavior started
in section \ref{Diag&Man}. 
Before proceeding further let us note that substituting
(\ref{alpha}) and (\ref{beta}) into (\ref{Vstar}) we can rewrite the
constant $v_*$ as
\begin{equation}
  v_* =1
  - \frac{r_0^2}{\xi + \eta}
  - \frac{s_0^2}{\xi \eta (\xi + \eta)}
  = -\Gamma_0.
%\label{V*1}
\end{equation}
It turns out that $v_*$ is nothing but? up to the sign< the quantity
$\Gamma_0$ introduced in section \ref{Diag&Man}. So, the constant $v_*$ indicates how
far the initial point is from the invariant manifold $\Gamma = 0$ at the initial moment.

Now let us consider the solution of the system in regions I
and II of the phase diagram. The common feature of these regions
is that the parameters $\xi$ and $\eta$ are complex: $\xi=\bar
\eta$, and the constants $\alpha$ and $\beta$ are related as
$\alpha=\bar \beta$.
In this case the solution of the system can be presented as
\begin{eqnarray}
  r & = & 2\frac{|\alpha|}{\sqrt{v}}\cos (\xi'' t +
  \phi_{\alpha}),
 \label{complexform1}
\\
   s & =& 2\frac{|\alpha||\xi|}{\sqrt{v}}\cos (\xi'' t +
  \phi_{\alpha}- \phi_{\xi})
   \label{complexform2}
\end{eqnarray}
and
\begin{equation}
  v=v_* e^{-2\xi' t}+2\frac{|\alpha|^2}{|\xi|}\cos (2\xi'' t +
  2\phi_{\alpha}- \phi_{\xi}) + 2\frac{|\alpha|^2}{\xi'},
  \label{complexV}
\end{equation}
where the following notations are introduced: $\xi = \xi' +
i\xi''$, $\phi_{\alpha}=\arg\alpha$, $\phi_\xi = \arg\xi$ .

Now it is easy to see that in the first region, where Re $\xi $
=  Re $\eta > 0$, the exponential term in $v$ vanishes as $t \to \infty$
for arbitrary initial conditions and the solution becomes
periodic which corresponds to the movement along the limit cycle.

\begin{figure}[h]%  
\begin{center}               
  \includegraphics[width=150mm]{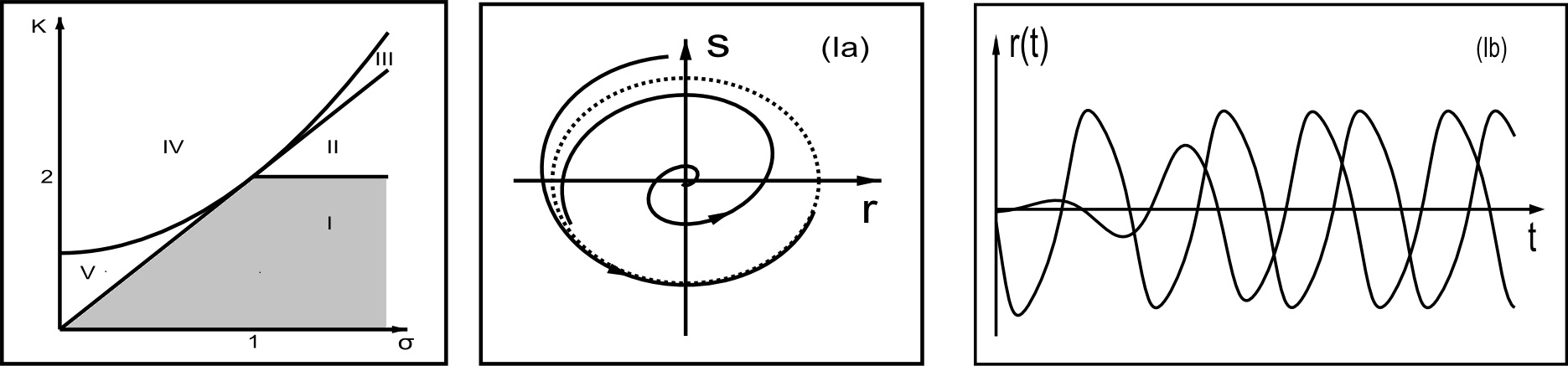}
\end{center}
    \caption{Typical phase trajectories (in the middle) and  the evolution of the centroid amplitudes 
 (on the right) corresponding to region I of the phase diagram (on the left) for different initial conditions.
  The limit cycle is plotted with dotted lines.}
  \label{fig:Fig2}
\end{figure}

\begin{figure}[h]%  
\begin{center}               
  \includegraphics[width=150mm]{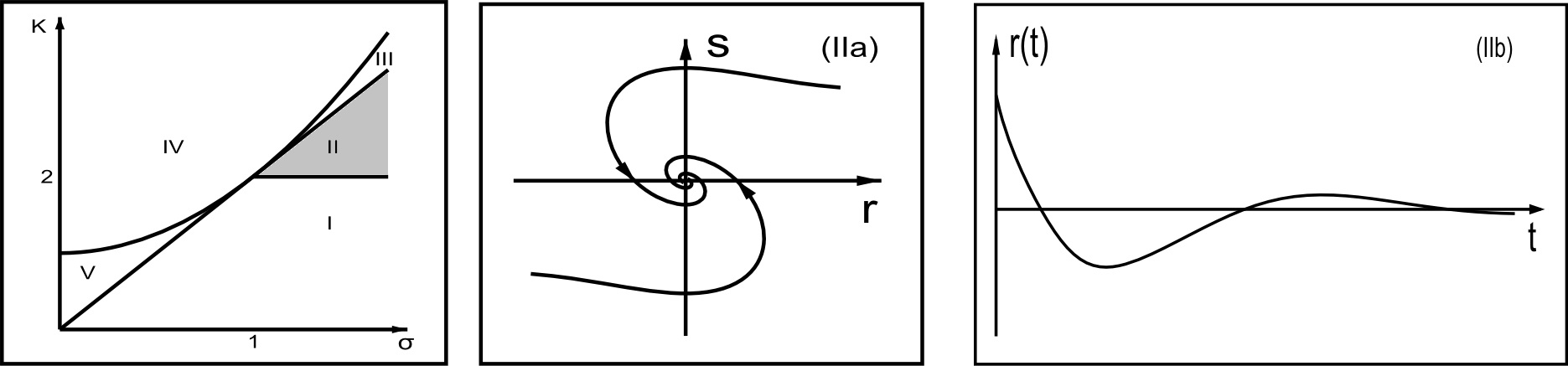}
\end{center}
    \caption{Typical phase trajectories and  the evolution of  of the centroid amplitude
   corresponding to region II of the phase diagram.}
  \label{fig:Fig3}
\end{figure}

\begin{figure}[h]%  
\begin{center}               
  \includegraphics[width=150mm]{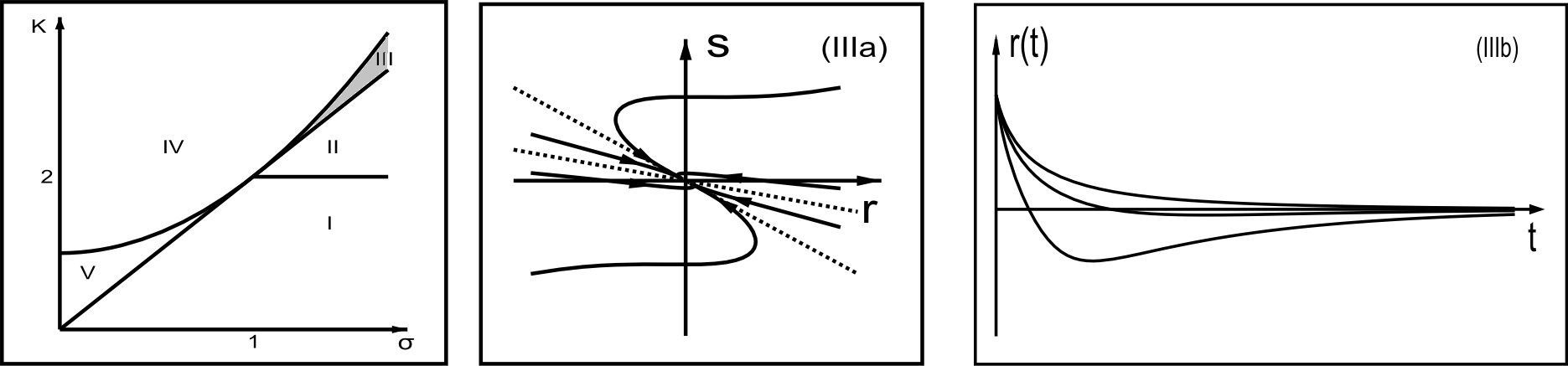}
\end{center}
    \caption{Typical phase trajectories and  the evolution of amplitudes of the centroid
  for region III of the phase diagram; the invariant manifolds are plotted with dotted lines.}
  \label{fig:Fig4}
\end{figure}

The movement along the limit cycle is described by
(\ref{complexform1}), (\ref{complexform2}) and (\ref{complexV})
with $v_*=-\Gamma_0=0$. The period of circulation along the cycle
is
\begin{equation}
  T=\frac{2\pi}{\xi''}.
  \label{period}
\end{equation}
The plots $r=r(t)$ in figure \ref{fig:Fig2} (Ib) present two typical solutions with
initial conditions corresponding to the points of the phase
portrait lying inside and outside the limit cycle. These solutions
fit the so-called `large oscillations' which were found
numerically in the previous works \cite{MetStr,DMO}.
For parameter values in region II Re $\xi $ =
Re $\eta < 0$, the exponent in (\ref{complexV}) tends to infinity
at $t \to \infty$ and our solutions, having a periodic
component, tends to the point of the stable focus (0,0) (figure \ref{fig:Fig3} (IIb)).
A similar asymptotic behavior is observed in 
case III: for all the initial conditions
$\lim\limits_{t\to\infty}r=\lim\limits_{t\to \infty}s=0$. The
difference is that now we do not have oscillations. This is a
manifestation of the fact that the stable point $(0,0)$ is the
stable node and not the focus as in case II (see Fig. \ref{fig:Fig4}). Both the
second and the third cases correspond to the amplitude death of
\cite{MetStr,DMO} when oscillations are quenched and the system evolves to
the stable equilibrium.

\begin{figure}[h]%  
\begin{center}               
  \includegraphics[width=150mm]{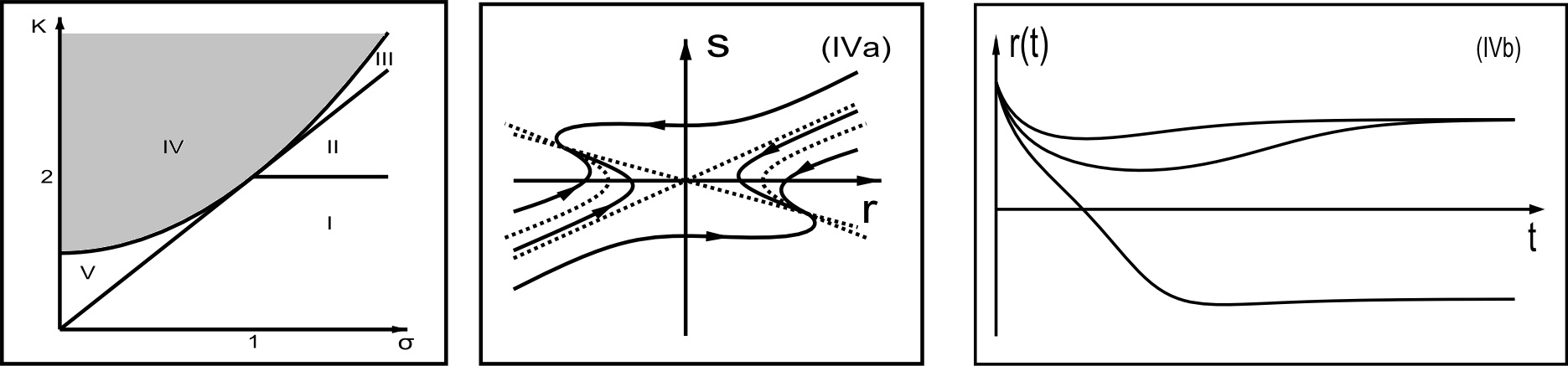}
\end{center}
    \caption{Typical phase trajectories and  the evolution of the centroid amplitudes
   for region IV of the phase diagram; the invariant manifolds are plotted with dotted lines.}
  \label{fig:Fig5}
\end{figure}

In the case IV expressions (\ref{solutionV})-(\ref{beta}) describe the
phase-locked solutions of our system when asymptotically the population as a
whole starts oscillate with the common amplitude $r^2=\xi$ (see figure \ref{fig:Fig5} (IVb)).

\begin{figure}[h]%  
\begin{center}               
  \includegraphics[width=150mm]{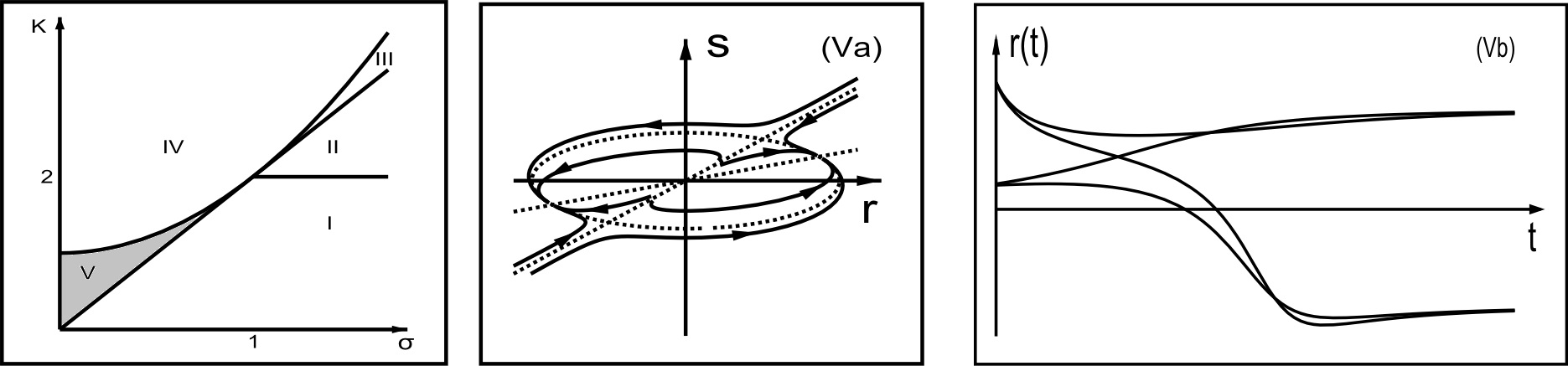}
\end{center}
    \caption{Phase portraits and typical evolution of amplitude
  $r=r(t)$  corresponding to region V of the phase diagram; the invariant manifolds are plotted with dotted lines.}
  \label{fig:Fig6}
\end{figure}

For the parameter values of region V of the phase
diagram the solutions of system (\ref{mainsystem}) are presented in figure \ref{fig:Fig6}. Though the asymptotic behavior of our system in the cases IV and V is visually similar these cases correspond to different geometry of their phase portraits which
can be essential in more elaborated descriptions. 

%
%%%%%%%%%%%%%%%%%%%%%%%%%%%%%%%%%%%%%%%%%%%%%%%%%%%%%%%%%%%%%%%%%%%%%%%%%%%%%%%%
\section{Conclusion.}
%%%%%%%%%%%%%%%%%%%%%%%%%%%%%%%%%%%%%%%%%%%%%%%%%%%%%%%%%%%%%%%%%%%%%%%%%%%%%%%%

In this paper we have integrate exactly the nonlinear system (\ref{mainsystem})  describing the evolution of macroscopic order parameters,  mean field (\ref{Z}) and  the amplitude-phase correlator (\ref{W}), of the ensemble of $N$ nonidentical globally coupled Stuart-Landau oscillators (\ref{Hforms}). Expressions (\ref{commonform1}), (\ref{commonform2}) and (\ref{solutionV})-(\ref{beta}) represent the complete analytical solution of system (\ref{system1}) on the two-dimensional  invariant manifold where the order parameters are orthogonal. We present a complete classification of
phase portraits and bifurcations, obtain explicit expressions for invariant manifolds
and derive analytical solutions for arbitrary initial data
and different regimes. The obtained results confirm the qualitative  and numerical description of (\ref{mainsystem}) offered in \cite{DMO} and lead us to expect
that analytical solutions can also be found in a less restricted situation, which will be the subject of future studies.

%%%%%%%%%%%%%%%%%%%%%%%%%%%%%%%%%%%%%%%%%%%%%%%%%%%%%%%%%%%%%%%%%%%%%%%%%%%%%%%
\section*{Acknowledgments.}

The authors are grateful to V~E~Vekslerchik for  fruitful discussions and his attention to this work.

%%%%%%%%%%%%%%%%%%%%%%%%%%%%%%%%%%%%%%%%%%%%%%%%%%%%%%%%%%%%%%%%%%%%%%%%%%%%%%%%

\end{document}